\documentclass [11pt]{article}

\usepackage{amsfonts,amssymb,amsmath}
\usepackage{epsfig,natbib,float}
\usepackage[ansinew]{inputenc}
\usepackage{graphicx}
\usepackage{booktabs}

\usepackage{fancyhdr}
\usepackage[dvipsnames]{xcolor}

\begin{document}
\
\begin{center}
\textbf{\large Partial Correlations in Compositional Data Analysis}

\bigskip
\textbf{Ionas Erb}$^{1}$\\
{\small
$^{1}$Centre for Genomic Regulation (CRG),\\ The Barcelona Institute of Science and Technology,\\
Barcelona, Spain; \textit{ionas.erb@crg.eu} \\
}
\end{center}

\bigskip
{\centerline{\bf Summary}}

\begin{quote}
Partial correlations quantify linear association between two variables adjusting for the influence of the remaining variables. They form the backbone for graphical models and are readily obtained from the inverse of the covariance matrix. For compositional data, the covariance structure is specified from log ratios of variables, so unless we try to ``open'' the data via a normalization, this implies changes in the definition and interpretation of partial correlations. In the present work, we elucidate how results derived by Aitchison (1986) lead to a natural definition of partial correlation that has a number of advantages over current measures of association. For this, we show that the residuals of log-ratios between a variable with a reference, when adjusting for all remaining variables including the reference, are reference-independent. Since the reference itself can be controlled for, correlations between residuals are defined for the variables directly without the necessity to recur to ratios except when specifying which variables are partialled out. Thus, perhaps surprisingly, partial correlations do not have the problems commonly found with measures of pairwise association on compositional data. They are well-defined between two variables, are properly scaled, and allow for negative association. By design, they are subcompositionally incoherent, but they share this property with conventional partial correlations (where results change when adjusting for the influence of fewer variables). We discuss the equivalence with normalization-based approaches whenever the normalizing variables are controlled for. We also discuss the partial variances and correlations we obtain from a previously studied data set of Roman glass cups.
\medskip

 \par

\end{quote}

\section{Introduction}

\subsection{Background and outline}
Since the publication of Aitchison's book on compositional analysis (Aitchison, 1986), awareness has increased about the fact that correlations between variables are problematic when these variables are parts of compositional data. To remedy the problems, a number of alternatives to correlation have been suggested. The most prominent is log-ratio variance (Aitchison, 1986), but there are also bounded (Aitchison, 2003) or scaled (Lovell et al., 2015) versions of it. All of these measures have their own problems: one would like to judge the value of log-ratio variance according to the intrinsic variability of its constituents, not only bound it between zero and one. Scalings however can lead to spurious results that become apparent when the underlying absolute data are known (Erb and Notredame, 2016). Apart from this, negative associations remain beyond the scope of all these measures of proportionality. Interestingly, the correlations between log-ratios that use the geometric mean as a reference, i.e., in clr-transformed data, have found little acceptance although they are perhaps the solution coming closest to a genuine correlation for compositional data. Its main drawback is subcompositional incoherence. However, the insistence on subcompositional coherence, and perhaps a certain reluctance to the use of correlation in the community, seem to have blocked the way towards embracing {\it partial} correlations as a valid way of analyzing pairwise association between variables summing to a constant. Clearly, here the coherence of results obtained on subsets of variables is neither required nor possible: all partial correlations change when removing variables we control for. To the best of our knowledge, this fact has not been exploited in compositional analysis, and work on partial correlation is lacking. This is especially surprising as the necessary mathematical results were derived by Aitchison in the 1980s already.\\ 
Indeed, when considering pairwise association between variables while adjusting for dependence on all other variables including the reference, the correlation that remains seems the best we can expect from a linear pairwise association measure of compositional parts. In this paper, we use some results from Aitchison (1986) to derive that correlating log-ratios when their reference variables are controlled for makes the correlations reference-independent in the sense that any reference from the variables we control for can be used interchangeably without altering the result. For a simple example, take the case of four compositional random variables $X_1,\dots,X_4$. Let us denote by
\begin{equation}
    r_{1,2|3,4}(X_1,\dots,X_4)=\mathrm{corr}\left(\log\frac{X_1}{X_4},\log\frac{X_2}{X_4}\left|\log\frac{X_3}{X_4}\right.\right)\label{first}
\end{equation}
the partial correlation between the logarithms of the first two variables referenced by $X_4$ adjusting for the influence of the log-ratio $X_3$ with the reference. The notation on the left-hand side seems ambiguous but it turns out that
\begin{multline}
    r_{1,2|3,4}(X_1,\dots,X_4)=\mathrm{corr}\left(\log\frac{X_1}{X_3},\log\frac{X_2}{X_3}\left|\log\frac{X_3}{X_4}\right.\right)\\
    =\mathrm{corr}\left(\log\frac{X_1}{\sqrt{X_3X_4}},\log\frac{X_2}{\sqrt{X_3X_4}}\left|\log\frac{X_3}{X_4}\right.\right)=\mathrm{corr}\left(\log\frac{X_1}{X_3},\log\frac{X_2}{X_4}\left|\log\frac{X_3}{X_4}\right.\right).
\end{multline}
Although partial correlations of the form (\ref{first}) are standard (in the sense that the theory of partial correlations applies directly to log ratios), it takes some results from the statistics of compositional data analysis to show that there is no dependence on the reference variable as long as it is included in the variables we control for. Thus partial correlations of log ratios are simpler than they appear at first sight, with some interesting implications. First, the number of correlations to interrogate is restricted to pairs and is thus drastically reduced with respect to the case where references have to be taken into account. Second, attempts to ``open" the data, i.e.\ to analyze them in absolute terms after normalization with an unchanged reference appear of little use if for most cases the same results can be derived without normalization. Third, the covariance matrix of the geometric-mean-referenced parts (i.e., of the clr-transformed data) can be used via its pseudoinverse to obtain all necessary results.

\subsection{Prerequisites}

\subsubsection{Data matrices as instances of compositional random vectors}\label{notprelim}
We start with a hypothetical, real-valued but positive $N\times D$ data matrix with elements $a_{ij}$, where samples are indexed in the rows and variables in the columns. The $N$ rows of this matrix can be considered instances of a $D$-dimensional random vector $A=(A_1,\dots,A_D)$. 
In order to avoid problems with zeros when dealing with log ratios, we also assume $A_j>0$ throughout. Let us now consider the closure of these instances, i.e.\ the data matrix $(x_{ij})$ resulting when we divide each element $a_{ij}$ by its row sum. The corresponding random vector $X=A/\sum(A_j)$ we call a compositional random vector. Compositional analysis concerns data where the total sum over a sample has no relevance to the analyst. This sum can vary between samples. The closure operation is just a convenient way of incorporating the fact that the total sum over the constituent random variables is arbitrary. 
The ``absolute" data $a_{ij}$ are considered unavailable here and were only mentioned to clarify that they can underlie $x_{ij}$. Data other than $a_{ij}$ can of course lead to the same $x_{ij}$. Put another way, we use the random vector $X$ as our representative for the equivalence class of random vectors with the same relative relationships between their random variables.\\
Let us now introduce yet another kind of random vector, $Z$, which we call the log-ratio transformed random vector. It is defined by $Z=\log(X/g(X))$, where $g$ denotes the geometric mean over the variables in the random vector. We can restrict the variables to an index set $\mathcal{A}$ and write this projection as $X_\mathcal{A}$. The resulting denominator $g(X_\mathcal{A})$ is called the reference. In the case of $\mathcal{A}$ containing all indices of the composition under consideration, the transformation is called the centered log-ratio transformation (clr), and the resulting random vectors we denote by $Z$, or $Z_\mathcal{A}$, whatever subset is considered. Throughout the manuscript, we let the set indexing $Z$ also indicate the variables over which the geometric mean is taken (unless a single variable is indicated, like for $Z_j$). In the case of the reference containing a single variable, usually $X_D$, the log-ratio transformation is called additive (alr). Note that the use of the index $D$ is a matter of convenience, and a variable of interest can just be moved to this last position. In general, results will depend on this choice. We will denote these alr-transformed random vectors by $Y$, and their restrictions to a subset $\mathcal{A}$ as $Y_\mathcal{A}$.\\
The two types of log-ratio transformed random vectors can be transformed into each other using a matrix designed for this purpose (Aitchison, 1986). Let us denote by $\mathbf{I}_d$ the $d$-dimensional identity matrix and by $\mathbf{j}_d$ the $d$-dimensional vector with entries 1. Let us define the matrix $\mathbf{F}=[\mathbf{I}_{D-1},-\mathbf{j}_{D-1}]$, i.e.\ $\mathbf{F}$ is the $(D-1)\times D$ matrix resulting from writing $\mathbf{I}_{D-1}$ and $-\mathbf{j}_{D-1}$ side by side. It is then easy to verify that we have
\begin{equation}
    Y=\mathbf{F}Z.\label{Z2Y}
\end{equation}

\subsubsection{Compositional covariance specifications and inverse variance}\label{mathprel}
Aitchison (1986) introduced three different matrices, each (equivalently) specifying the full covariance structure of a compositional data set. The first matrix has the log-ratio variances $\mathrm{var}(\log(X_i/X_j))$ as elements and will not concern us further. The other two specifications are covariance matrices of the log-ratio transformed random vectors. With the notation from the previous section, the $(D-1)\times(D-1)$ matrix of alr-transformed random vectors is $\mathbf{\Sigma}=(\sigma_{ij})=\left(\mathrm{cov}(Y_i,Y_j)\right)=\mathrm{var}(Y)$.  Finally, the third matrix is the covariance of clr-transformed random vectors $\mathbf{\Gamma}=(\gamma_{ij})=\left(\mathrm{cov}(Z_i,Z_j)\right)=\mathrm{var}(Z)$. The change in log-ratio transformation results in a singular $D\times D$ matrix, with the singularity resulting from the constraint $\sum_jZ_j=0$. Similarly to (\ref{Z2Y}), we can also transform the covariance matrices into each other:
\begin{equation}
    \mathbf{\Sigma}=\mathbf{F}\mathbf{\Gamma}\mathbf{F}^T.\label{SigmaGamma}    
\end{equation}
With partial correlations in mind, an important result derived by Aitchison (1986) concerns the relationship between the pseudoinverse $\mathbf{\Gamma^-}$ of $\mathbf{\Gamma}$ and the inverse variance $\mathbf{\Sigma}^{-1}$. We have
\begin{equation}
    \mathbf{\Gamma^-}=\mathbf{F}^T\mathbf{\Sigma}^{-1}\mathbf{F}\label{pseudo}
\end{equation}
(see Property 5.6 (a) in Aitchison, 1986). In this article, we will essentially elucidate the implications of this relationship.

\subsubsection{Partial correlation: The standard setting}
Due to its unconstrained nature, the covariance matrix $\mathbf{\Sigma}$ can be used to obtain partial correlations (between log ratios having $X_D$ as a reference) in the standard way. Let us quickly review this procedure (see, e.g., Whittaker, 1990). In the following, to achieve more economical expressions, we assume that all log-ratio transformed random variables are centered, i.e.\ their averages are zero (in case they are not, we just have to subtract their average from them). Let the random vector $Y_\mathcal{C}$ be composed of the set of random variables having indices in $\mathcal{C}$, i.e.\ $(Y_i)_{i\in\mathcal{C}}$. The linear least squares predictor (LLSP) of $Y_j$ given $Y_\mathcal{C}$ is then defined by
\begin{equation}
    \hat{Y}_j(Y_\mathcal{C})=\mathrm{cov}(Y_j,Y_\mathcal{C})\mathrm{var}(Y_\mathcal{C})^{-1}Y_\mathcal{C}.\label{llsp}
\end{equation}
Here, $\mathrm{cov}(Y_j,Y_\mathcal{C})$ is the row vector containing the covariances of $Y_j$ with the scalar random variables in $Y_{\mathcal{C}}$, and $\mathrm{var}(Y_\mathcal{C})^{-1}$ is the inverse covariance matrix of these random variables. Note that, as a consequence of this definition, $\hat{Y}_j$ lies in the space spanned by $Y_\mathcal{C}$ and the residual $Y_j-\hat{Y}_j(Y_\mathcal{C})$ is orthogonal to that space.\footnote{More precisely, the predictor and the residual generate $N$-sample vectors that lie in/are orthogonal to the space spanned by the $N$-sample vectors generated from the random variables in $Y_\mathcal{C}$.} The residual variance
\begin{equation}
    \mathrm{var}(Y_j|Y_\mathcal{C})=\mathrm{var}\left(Y_j-\hat{Y}_j(Y_\mathcal{C})\right)
\end{equation}
is a useful summary statistic that tells us how well $Y_j$ can be predicted from the variables $Y_\mathcal{C}$. It is also known as partial variance. More generally, the partial covariance between two scalar random variables is defined as the covariance between their residuals:
\begin{equation}
    \mathrm{cov}(Y_i,Y_j|Y_\mathcal{C})=\mathrm{cov}\left(Y_i-\hat{Y}_i(Y_\mathcal{C}),Y_j-\hat{Y}_j(Y_\mathcal{C})\right).
\end{equation}
Partial correlations are obtained scaling by the corresponding residual variances in the standard way. There is, however, a shortcut to this. An important result states that partial variances and correlations adjusting for all the remaining variables can be obtained from the inverse covariance matrix. We have 
\begin{eqnarray}
    \mathrm{var}(Y_j|Y_{\{1,\dots,D-1\}\backslash j})&=&1/\sigma^{(-1)}_{jj},\label{partvar}\\
    \mathrm{corr}(Y_i,Y_j|Y_{\{1,\dots,D-1\}\backslash\{i,j\}})&=&\frac{-\sigma^{(-1)}_{ij}}{\sqrt{\sigma^{(-1)}_{ii}\sigma^{(-1)}_{jj}}},\label{partcorr}
\end{eqnarray}
where $\sigma^{(-1)}_{ij}$ denote the elements of $\mathbf{\Sigma}^{-1}$.

\section{Partial correlation on compositional data} 
\subsection{The residual is independent of the choice of alr transformation}
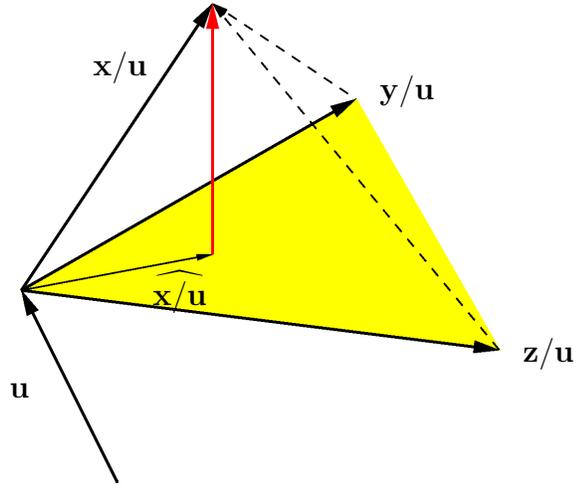
\begin{figure}
\centering
\begin{picture}(0,0)%
\includegraphics{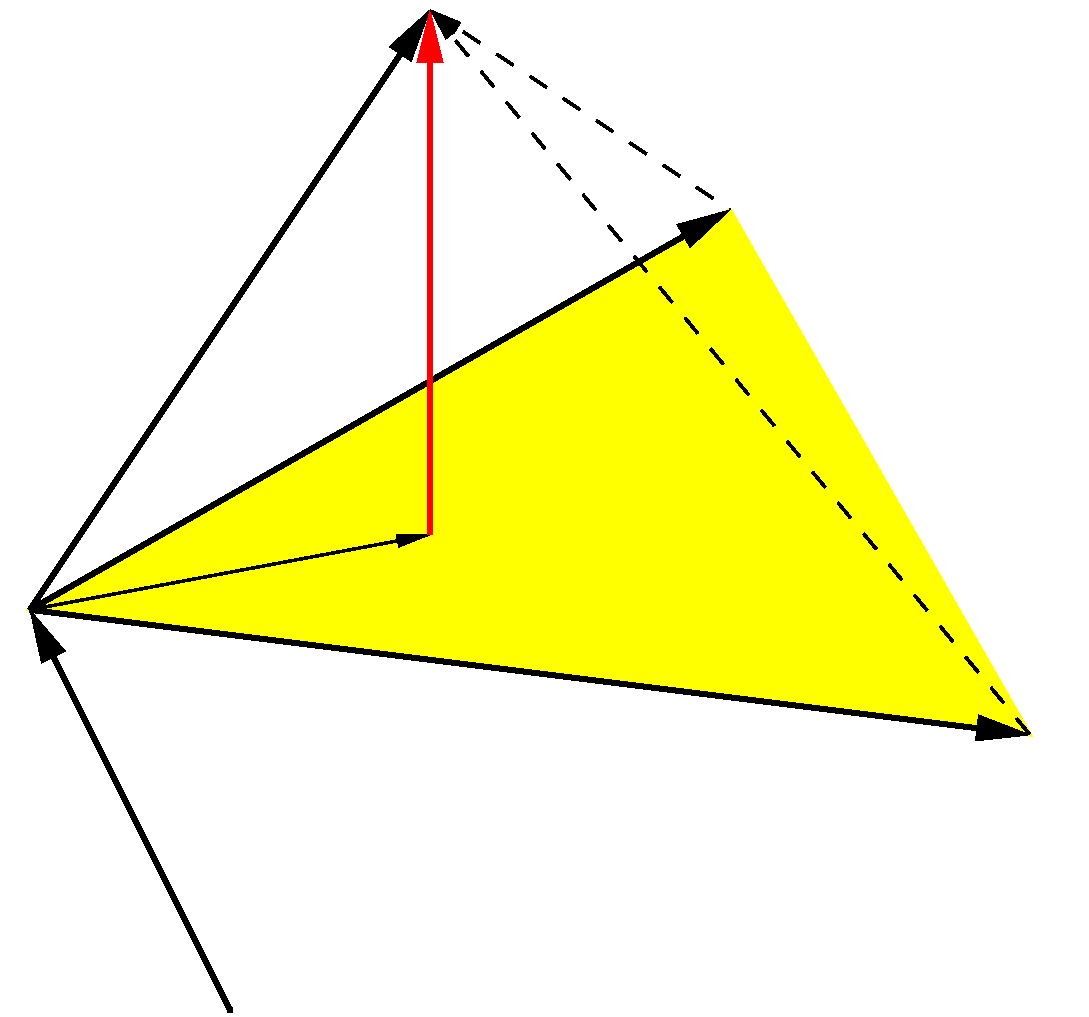}%
\end{picture}%
\setlength{\unitlength}{3947sp}%
\begingroup\makeatletter\ifx\SetFigFont\undefined%
\gdef\SetFigFont#1#2#3#4#5{%
  \reset@font\fontsize{#1}{#2pt}%
  \fontfamily{#3}\fontseries{#4}\fontshape{#5}%
  \selectfont}%
\fi\endgroup%
\begin{picture}(3255,3044)(4711,-4583)
\put(4726,-4036){\makebox(0,0)[lb]{\smash{{\SetFigFont{12}{14.4}{\rmdefault}{\mddefault}{\updefault}{$\mathbf{u}$}%
}}}}
\put(5251,-2011){\makebox(0,0)[lb]{\smash{{\SetFigFont{12}{14.4}{\rmdefault}{\mddefault}{\updefault}{$\mathbf{x}/\mathbf{u}$}%
}}}}
\put(7951,-3811){\makebox(0,0)[lb]{\smash{{\SetFigFont{12}{14.4}{\rmdefault}{\mddefault}{\updefault}{$\mathbf{z}/\mathbf{u}$}%
}}}}
\put(5626,-3450){\makebox(0,0)[lb]{\smash{{\SetFigFont{12}{14.4}{\rmdefault}{\mddefault}{\updefault}{$\widehat{\mathbf{x}/\mathbf{u}}$}%
}}}}
\put(7051,-2161){\makebox(0,0)[lb]{\smash{{\SetFigFont{12}{14.4}{\rmdefault}{\mddefault}{\updefault}{$\mathbf{y}/\mathbf{u}$}%
}}}}
\end{picture}%
\caption{The residual of $\mathbf{x}$ is independent of its reference. Shown is the $D=4$ case with (logged) vectors $\mathbf{x}$, $\mathbf{y}$, $\mathbf{z}$ and $\mathbf{u}$, the latter serving as a reference. To simplify, logs are omitted in the vector annotations. The linear least squares predictor of $\log(\mathbf{x}/\mathbf{u})$ with respect to $\left(\log(\mathbf{y}/\mathbf{u}),\log(\mathbf{z}/\mathbf{u})\right)$ is shown as the thin arrow lying in the yellow plane. The residual vector is shown in red. It can be seen that replacing $\log(\mathbf{x}/\mathbf{u})$ by $\log(\mathbf{x}/\mathbf{y})$ or $\log(\mathbf{x}/\mathbf{z})$ (dashed lines) results in the same residual.}
\label{residualfig}
\end{figure}

While $\mathbf{\Sigma}$ is a standard covariance matrix, in a compositional context its usefulness may not be apparent as it depends on the reference chosen. Rather surprisingly, it turns out that the inverse $\mathbf{\Sigma}^{-1}$ is essentially independent of the choice of reference. This is implied by (\ref{pseudo}), but although the equation may seem unsurprising given (\ref{SigmaGamma}), the fact that a matrix containing all the parts can be obtained from one where a part is sacrificed as reference appears somewhat mysterious. This can be understood better when looking from another angle. Remember that inverse covariance matrices have diagonal elements related to the variance of the residuals, see (\ref{partvar}). The fact that for $\mathbf{\Sigma}^{-1}$ the choice of reference has little importance can be made sense of if the residuals themselves are independent of this choice.\\
Remember that $Y_j$ is an alr-transformed random variable with reference $X_D$ and that we have $\mathbf{\Sigma}=\mathrm{var}(Y)$ with $Y$ having $D-1$ components. For removing $Y_j$ form $Y$, let us define the index set $\mathcal{C}_j=\{1,\dots,D-1\}\backslash j$. We now show that the residual 
\begin{equation}
   Y_j-\hat{Y}_j\left(Y_{\mathcal{C}_j}\right)\label{residual}
\end{equation}
remains the same irrespective of which reference is chosen from the set $\{X_k|k=1,\dots,D,k\ne j\}$ for constructing its constituent random variables $\log X_i/X_k$, $i\ne j,k$. For this, we first have to clarify that the space generated from the explanatory log ratios (the random vector $Y_{\mathcal{C}_j}$ we are adjusting for) remains unchanged when choosing another variable for reference. Intuitively, a change of reference in the variables we adjust for corresponds to an overall subtraction of one of these variables (the one that has the new reference in the numerator) and neither reduces nor moves us outside the space generated by $Y_{\mathcal{C}_j}$. As prediction is equivalent with linear projection into this space, the LLSP will remain the same. A formal proof that permuting the variables in $Y_{\mathcal{C}_j}$ does not change the LLSP is presented in the Appendix.\footnote{Note that this does not imply that the LLSP is independent of the choice of reference, i.e., in general we have $\widehat{\log(X_j/X_D)}\ne\widehat{\log(X_j/X_k)}$.}
Now let us come back to the residual. A change from reference $X_D$ to reference $X_k$ is achieved by subtracting $Y_k$ from $Y_j$ ($k\ne j,D$):
\begin{equation}
    Y_j-Y_k-(\widehat{Y_j-Y_k})=Y_j-\hat{Y}_j-(Y_k-\hat{Y}_k),\label{changeref}
\end{equation}
where the equality comes form the linearity of the predictor. With the definition of the LLSP (\ref{llsp}) and denoting the elements of $\mathrm{var}(Y_{\mathcal{C}_j})^{-1}$ by $c^{(-1)}_{il}$ we have
\begin{equation}
    \hat{Y}_k\left(Y_{\mathcal{C}_j}\right)=\mathrm{cov}(Y_k,Y_{\mathcal{C}_j})\mathrm{var}(Y_{\mathcal{C}_j})^{-1}Y_{\mathcal{C}_j}=\left(\sum_{i\in\mathcal{C}_j}\sigma_{ki}c^{(-1)}_{il}\right)^T_{l\in\mathcal{C}_j}Y_{\mathcal{C}_j}=(\delta_{kl})^T_{l\in\mathcal{C}_j}Y_{\mathcal{C}_j}\\=Y_k.
\end{equation}
Here, $\delta$ denotes the Kronecker delta. It appears because the elements of $\mathbf{\Sigma}$ and of $\mathrm{var}(Y_{\mathcal{C}_j})$ coincide for indices in $\mathcal{C}_j$. Comparing with the right-hand side of (\ref{changeref}), we see that the term in brackets vanishes, thus establishing the independence of the residual from a reference $X_k$, $k\ne j$. Figure \ref{residualfig} shows an intuitive rendering of the argument we provided. For the argument it is crucial that the reference is coming from the constituent variables of the ratios we adjust for.

\subsection{Connection with clr transformation}
We have seen that our residuals do not depend on the particular alr transformation chosen (as long as the reference occurs as a variable in the ratios we are adjusting for). What about the reference coming from a clr transformation? Note that for an equivalent formulation, we need to consider the clr on the subspace of variables we are adjusting for. A similar argument as in the previous section (see Appendix) shows that an equivalent expression of our residual (\ref{residual}) that is symmetric with respect to the reference variables is
\begin{equation}
    Z_j-\hat{Z}_j\left(Z_{\mathcal{D}_j}\right),\label{residual2}
\end{equation}
where the geometric mean reference of $Z_j$ goes over the indices in $\mathcal{D}_j=\{1,\dots,D\}\backslash j$.
Now from (\ref{pseudo}) we know that the diagonal elements of the pseudoinverse of $\mathbf{\Gamma}$ are
\begin{equation}
    \gamma_{jj}^{(-)}=\sigma_{jj}^{(-1)},j=1,\dots,D-1.
\end{equation}
Due to the symmetry of $\mathbf{\Gamma}$ with respect to reference, this also has to hold after permuting labels when constructing $\mathbf{\Sigma}$ (see Appendix for a proof). Thus, with (\ref{partvar}), we conclude that for all indices $j=1,\dots,D$, the residual variance can be calculated from
\begin{equation}
    \mathrm{var}\left(Z_j\left|Z_{\mathcal{D}_j}\right.\right)=1/\gamma_{jj}^{(-)}.\label{partvar_gen}
\end{equation}
Similarly, the off-diagonal elements of $\mathbf{\Gamma}^-$ coincide with the ones of $\mathbf{\Sigma}^{-1}$ when they have indices $i,j<D$. Again invoking invariance with respect to label permutations, we conclude that for all $i,j=1,\dots,D$, partial correlations can be calculated from
\begin{equation}
   \mathrm{corr}\left(Z_i,Z_j\left|Z_{\mathcal{D}_{ij}}\right.\right)=\frac{-\gamma^{(-)}_{ij}}{\sqrt{\gamma^{(-)}_{ii}\gamma^{(-)}_{jj}}},\label{partcorr_gen}
\end{equation}
where we used the notation $\mathcal{D}_{ij}=\{1,\dots,D\}\backslash\{i,j\}$. Note that here the $Z_i$, $Z_j$ are different from the ones in (\ref{partvar_gen}) because the geometric mean reference is taken over $X_{\mathcal{D}_{ij}}$, not $X_{\mathcal{D}_j}$. 

\subsection{A general expression, correlation matrices, $R^2$}
To summarize, we have shown that we can use as a reference either variable occurring in the explanatory ratios or the geometric mean taken over them. For compositional data it is thus unambiguous to define partial variances and correlations directly for the parts $X_i$ of the composition via
\begin{eqnarray}
    \sigma^2_{j|\mathcal{D}_j}(X)&=&1/\gamma_{jj}^{(-)},\label{partvar_parts}\\
    r_{ij|\mathcal{D}_{ij}}(X)&=&\frac{-\gamma^{(-)}_{ij}}{\sqrt{\gamma^{(-)}_{ii}\gamma^{(-)}_{jj}}},\label{partcorr_parts}
\end{eqnarray}
where we mean the variances and correlations evaluated on residuals of the form (\ref{residual}) or (\ref{residual2}) and with any reference coming from the log ratios we are adjusting for.\\
Correlation matrices are easier to interpret than covariance matrices, and the inverses of correlation matrices are often used in practice. Although $\mathbf{\Sigma}^{-1}$ and $\mathbf{\Gamma}^-$ coincide where they can, their correlation counterparts do not. Indeed, when scaling $\mathbf{\Sigma}$ and $\mathbf{\Gamma}$ to have unit entries in the diagonal,\footnote{Transforming a covariance into a correlation matrix is achieved by multiplying from the left and the right a suitably scaled diagonal matrix.} after inverting the scaled version of $\mathbf{\Sigma}$ and pseudo-inverting the scaled version of $\mathbf{\Gamma}$, we have to re-scale both to unit diagonal to make them coincide again. Their multiple correlation coefficients $R^2$, however, remain different. As they quantify the amount of variance of $Y_j$ or $Z_j$ explained by their respective predictors (which depend on the reference), they have to be reference-dependent:  
\begin{eqnarray}
    \mathrm{var}(\hat{Y}_j)/\mathrm{var}(Y_j)&=&1-1/\left(\sigma_{jj}\sigma_{jj}^{(-1)}\right),\\
    \mathrm{var}(\hat{Z}_j)/\mathrm{var}(Z_j)&=&1-1/\left(\gamma_{jj}\gamma_{jj}^{(-)}\right)\label{symmRsq}.
\end{eqnarray}
Although $Y_j$ and $Z_j$ have the same partial variance $\sigma^2_{j|\mathcal{D}_j}(X)=1/\gamma_{jj}^{(-)}=1/\sigma_{jj}^{(-1)}$, their zero-order variances are different. When an $R^2$ needs to be reported, (\ref{symmRsq}) would be a likely candidate or otherwise one would have to have a good reason to choose a more specific reference $X_D$. 

\subsection{Comparison with normalization-based partial correlations}\label{normsection}
An approach that is sometimes pursued (e.g., in genomics) consists in trying to recover the underlying absolute data $a_{ij}$ mentioned in section \ref{notprelim}. This is achieved by means of a specific normalization, i.e., by multiplication of the rows of $(x_{ij})$ with factors $s_i$ that are proportional to $\sum_ja_{ij}$. These factors are unknown in principle (they got lost by the closure operation). They can sometimes be inferred from assumptions concerning the knowledge of (on average) unchanged $a_{ij}$ across rows (see the supplement to Quinn et al., 2018). If the assumptions are fulfilled, the resulting data matrix is no longer relative and there is no need for analyzing ratios. Note however that we can write the normalized data as a special case of log-ratio transformed data $Z=\log(X/g(X_\mathcal{U}))$, where the set $\mathcal{U}$ contains only indices of variables for which we have $g(A_\mathcal{U})=$const. (We have just transformed to logarithms of the normalized data, a common procedure in genomics.) When comparing with the partial correlations for log-ratio transformed data with more general references (\ref{partcorr_parts}), we see that they coincide (and can be obtained assumption-free, without normalization) as long as the log ratios we are adjusting for also contain the variables indexed by $\mathcal{U}$. 

\section{Application to Roman glass-cup data}

 We evaluate (\ref{partvar_parts}) and (\ref{partcorr_parts}) on a data set of 11 oxides and elements composing the glass of 47 Roman cups excavated in Colchester (Cool and Price, 1994). We use the version of the data presented in (Baxter et al., 1990). The few previous analyses have focused on ordination. One of the conclusions of Greenacre's recent analysis (Greenacre, 2018) is that the ratio of SiO$_2$/CaO is the significant dimension of variation, followed by the ratios of SiO$_2$/Sb and Na$_2$O/Sb, which were found to be of lesser influence for the multivariate structure. In Table \ref{pvartab}, beside the average weight percentages of the various oxides and elements, we show the partial variances and log-ratio variances (with respect to the geometric mean). For better readability, both of them are divided by total variance as obtained from the trace of $\mathbf{\Gamma}$. Then, $R^2$ with the mean reference is shown. In the last two columns we also present log-ratio variance and $R^2$ with respect to SiO$_2$. The oxides occurring in high weight percentages, with the exception of SiO$_2$, tend to have low variances. The relatively high variance of SiO$_2$ is however well explained by the other variables (highest $R^2$ with geometric mean reference). This may explain the good properties as a reference reported by Greenacre (2018). The two variables with the highest residual variance (MnO and Sb) also contribute most variance to the data. The values of $R^2$ with respect to SiO$_2$ show that Fe$_2$O$_3$ and Al$_2$O$_3$ are best predicted using this reference.
 \begin{table}[ht]
 \caption{Single-variable quantities for Roman glass cups. All values are given in percent. }
\begin{center}
\begin{tabular}{*7c} 
\hline
 oxide, element & av.\ weight & res.\ var/tot.\ & var/tot.\ (mean) & $R^2$ (mean) & var/tot.\ (SiO$_2$) & $R^2$ (SiO$_2$)\\
\hline
SiO$_2$ & 72 & 0.86 & 7.7 & 89 & - & -\\
Al$_2$O$_3$ & 1.9 & 0.37 & 2.5 & 85 & 2.0 & 90\\
Na$_2$O & 18 & 0.75 & 3.6 & 79 & 1.4 & 70\\
Fe$_2$O$_3$ & 0.30 & 1.5 & 5.9 & 75 & 11 & 93\\
MgO & 0.46 & 3.4 & 6.4 & 46 & 8.3 & 78\\
CaO & 5.6 & 1.2 & 1.9 & 39 & 3.5 & 81\\
TiO$_2$ & 0.07 & 3.5 & 4.9 & 29 & 8.2 & 77\\
MnO & 0.01 & 21 & 29 & 26 & 30 & 61\\
Sb & 0.35 & 22 & 28 & 23 & 22 & 45\\
P$_2$O$_5$ & 0.05 & 4.3 & 5.3 & 19 & 8.0 & 70\\
K$_2$O & 0.49 & 3.6 & 4.2 & 15 & 5.9 & 67\\
\hline
\end{tabular}
\end{center}
\label{pvartab}
\end{table}
 In table \ref{pcorrtab} we show the five strongest partial correlations as well as a plug-in estimate of their local false discovery rates obtained under permutations of the samples for each variable in the data matrix (see Appendix). Of note, permuting the residuals directly leads to a less severe test and much lower $q$-values, presumably because a natural dependence among the residuals gets destroyed.  
 \begin{table}[ht]
 \caption{Partial correlations and their $q$-values of five pairs of variables}
\begin{center}
\begin{tabular}{*4c} 
\hline
 variable 1 & variable 2 & partial correlation & q-value\\
\hline
Fe$_2$O$_3$ & Al$_2$O$_3$ & 0.73 & 0.03\\
Al$_2$O$_3$ & SiO$_2$ & 0.72 & 0.03\\
Fe$_2$O$_3$ & SiO$_2$ & -0.66 & $<10^{-4}$\\
Na$_2$O & CaO & 0.60 & 0.18\\
MgO & Fe$_2$O$_3$ & 0.43 & 0.46\\
\hline
\end{tabular}
\end{center}
\label{pcorrtab}
\end{table}
While Greenacre's analysis stresses the importance of SiO$_2$ and the oxides with high weight percentages for the multivariate structure, partial correlations and $R^2$ seem to point us to the importance of Fe$_2$O$_3$ and Al$_2$O$_3$ for the understanding of mutual dependencies. Of note is the conspicuous negative correlation between SiO$_2$ and Fe$_2$O$_3$, the only negative association of importance that we find in this data set. It would remain undetected with current association measures.   

\section{Discussion} 
A proposal for partial proportionality by Erb and Notredame (2016) only controls for one ratio and thus introduces an additional parameter. Already the reference can be considered a nuisance parameter and needs justification when discussing results. The natural definition of partial correlation elucidated in this paper treats the reference like just another variable that is controlled for. While we cannot quite get around discussing its choice for $R^2$, partial variances and partial correlations are defined reference-free and can be evaluated as efficiently as in absolute analysis. Application to graphical models is a natural next step. For applications in genomics, e.g., the equivalence of the compositional approach with the one using a normalization of the data is interesting. The equivalence holds as long as the normalizing variables are controlled for in the sense discussed in section \ref{normsection}. In genomics it will also be necessary to modify the analysis for the many-variables setting. Here, the empirical covariance matrix is a poor estimator that can be improved by regularization, also allowing for the necessary covariance inversion. 

\section*{Acknowledgements}
I thank Cedric Notredame and Thom Quinn for their support and encouragement.

\section*{References}

\hangindent=0.5cm \hangafter=1 %
\hspace{\parindent}Aitchison, J (1986).
\newblock {\em The Statistical Analysis of Compositional Data.}
\newblock London: Chapman \& Hall.

\hangindent=0.5cm \hangafter=1 %
Aitchison, J (2003).
\newblock {\em A concise guide to compositional data analysis.}
\newblock {\em The 2nd Compositional Data Analysis Workshop; Girona, Spain.}

\hangindent=0.5cm \hangafter=1 %
Baxter MJ, Cool HEM, Heyworth MP (1990).
\newblock Principal component and correspondence analysis of compositional data: some similarities. 
\newblock {\em J Appl Stat}~{\em 17}, pp 229--235.

\hangindent=0.5cm \hangafter=1 %
Cool, HEM and Price, J (Eds.) (1994).
\newblock {\em Colchester Archeological Report 8: Roman vessel glass from excavations in Colchester, 1971-85.}
\newblock Hunstanton: Witley. 

\hangindent=0.5cm \hangafter=1 %
Erb, I. and Notredame, C (2016).
\newblock How should we measure proportionality on relative gene expression data?
\newblock {\em Theory in Biosciences}~{\em 135\/}(1-2), pp 21--36.

\hangindent=0.5cm \hangafter=1 %
Greenacre, M (2018). 
\newblock Variable Selection in Compositional Data Analysis Using Pairwise Logratios. 
\newblock {\em Mathematical Geosciences} (Online First), pp 1--34.

Hastie, T, Tibshirani, R and Friedman, J (2001)
\newblock {\em The elements of statistical learning.} 
New York: Springer.

\hangindent=0.5cm \hangafter=1 %
Lovell, D, Pawlowsky-Glahn, V, Egozcue, J, Marguerat, S and B{\"a}hler, J (2015).
\newblock Proportionality: a valid alternative to correlation for relative data.
\newblock {\em PLoS Computational Biology},~{\em 11\/}(3).

\hangindent=0.5cm \hangafter=1 %
Quinn, TP, Erb, I, Richardson, MF, Crowley, TM (2018).
\newblock {\em Understanding sequencing data as compositions: an outlook and review},~{\em34\/}(16), pp 2870--2878.

\hangindent=0.5cm \hangafter=1
Whittaker, J. (1990).
\newblock {\em Graphical models in applied multivariate statistics}.
\newblock Chichester: Wiley.

\section*{Appendix}
\subsection*{Independence of LLSP from permutation of parts in the explanatory ratios}
Let $\mathbf{P}$ be a $D\times D$ permutation matrix (obtained from permuting the rows of the identity matrix). When applying $\mathbf{P}$ to $X$, it changes the order of the random variables in $X$. Together with our definition of $\mathbf{F}$ from section \ref{mathprel}, we define
\begin{equation}
    \mathbf{Q}_P=\mathbf{F}\mathbf{P}\mathbf{F}^T\mathbf{H}^{-1},
\end{equation}
where $\mathbf{H}$ is defined as the $(D-1)\times(D-1)$ matrix resulting from adding a matrix of units to the identity matrix.\footnote{Note that we have $Z=\mathbf{F}^T\mathbf{H}^{-1}Y$. The matrix $\mathbf{F}^T\mathbf{H}^{-1}$ is a right inverse of $\mathbf{F}$.}  Also, it can be obtained from $\mathbf{H}=\mathbf{F}\mathbf{F}^T$. Note that, when applying $\mathbf{Q}_P$ to $Y$, it yields a permutation of parts in the constituent variables, e.g., for a suitable $\mathbf{P}$, $Y=\left(\log(X_1/X_3),\log(X_2/X_3)\right)$ becomes $\left(\log(X_1/X_2),\log(X_3/X_2)\right)=Y^P$. More formally, Property 5.2 (b) in (Aitchison, 1986) states that
\begin{eqnarray}
    Y^P&=&\mathbf{Q}_PY,\label{iv}\\
    \mathbf{\Sigma}_P&=&\mathbf{Q}_P\mathbf{\Sigma}\mathbf{Q}_P^T.\label{v}
\end{eqnarray}
Here, $\mathbf{\Sigma}$ and $\mathbf{\Sigma}_P$ denote the covariance matrices of $Y$ and $Y_P$, respectively. Let us now use versions of these matrices to contain only variables with indices from $\mathcal{C}_j=\{1,\dots,D-1\}\backslash j$. For ease of notation, we will not change the notation for $\mathbf{Q}_P$ for the lower-order version, but to use the same notation as in the main text, let us use the shorthand $\mathbf{C}=\mathrm{var}(Y_{\mathcal{C}_j})$ instead of $\mathbf{\Sigma}$. (Remember that in $Y_{\mathcal{C}_j}$, $X_D$ occurs in the denominator of the ratios.) We can now show that $\hat{Y_j}(Y^P_{\mathcal{C}_j})=\hat{Y}_j(Y_{\mathcal{C}_j})$:
\begin{multline}
    \hat{Y_j}(Y^P_{\mathcal{C}_j})=\mathrm{cov}\left(Y_j,Y^P_{\mathcal{C}_j}\right)\mathrm{var}(Y^P_{\mathcal{C}_j})^{-1}Y^P_{\mathcal{C}_j}=\mathrm{cov}\left(Y_j,\mathbf{Q}_PY_{\mathcal{C}_j}\right)\mathbf{C}_P^{-1}\mathbf{Q}_PY_{\mathcal{C}_j}\\
    =\mathrm{cov}\left(Y_j,Y_{\mathcal{C}_j}\right)\mathbf{Q}_P^T\mathbf{C}_P^{-1}\mathbf{Q}_PY_{\mathcal{C}_j}=\mathrm{cov}\left(Y_j,Y_{\mathcal{C}_j}\right)\mathbf{Q}_P^T\left(\mathbf{Q}_P\mathbf{C}\mathbf{Q}_P^T\right)^{-1}\mathbf{Q}_PY_{\mathcal{C}_j}\\
    =\mathrm{cov}\left(Y_j,Y_{\mathcal{C}_j}\right)\mathbf{Q}_P^T\left(\mathbf{Q}_P^T\right)^{-1}\mathbf{C}^{-1}\mathbf{Q}_P^{-1}\mathbf{Q}_PY_{\mathcal{C}_j}=\mathrm{cov}\left(Y_j,Y_{\mathcal{C}_j}\right)\mathbf{C}^{-1}Y_{\mathcal{C}_j}=\hat{Y_j}(Y_{\mathcal{C}_j}).
\end{multline}
Here, the first line uses the definition of the LLSP (\ref{llsp}), the identity (\ref{iv}) and the definition of $\mathbf{C}_P$. The first equality of the second line comes from the bilinearity of the covariance, c.f.\ Proposition 5.1.2 in (Whittaker, 1990). The next equality follows from (\ref{v}). Some simple matrix identities and again the definition of the LLSP conclude the proof. 

\subsection*{Equivalent formulation of residual using clr transformation}
Let us start with the explanatory ratios again i.e.\ let us show that $\hat {Y}_j(Y)=\hat {Y}_j(Z)$. (For simplicity, we partial on the entire vector $Y$, but the same argument holds for some $Y_\mathcal{C}$.)
\begin{multline}
    \hat {Y}_j(Y)=\mathrm{cov}(Y_j,Y)\mathbf{\Sigma}^{-1}Y=\mathrm{cov}(Y_j,\mathbf{F}Z)\mathbf{\Sigma}^{-1}\mathbf{F}Z=\mathrm{cov}(Y_j,Z)\mathbf{F}^T\mathbf{\Sigma}^{-1}\mathbf{F}Z=\mathrm{cov}(Y_j,Z)\mathbf{\Gamma}^{-}Z=\hat{Y}_j(Z).\label{YisZ}
\end{multline}
For the first four equalities we were using (\ref{llsp}), (\ref{Z2Y}), the bilinearity of covariance (see previous proof) and (\ref{pseudo}). The last equality we can consider a definition (as in the original definition of the LLSP the covariance matrix needs to be invertible and is here replaced by the pseudoinverse). As this argument was independent of the variable we are predicting, we also conclude that $\hat {Z}_j(Y)=\hat {Z}_j(Z)$.\\
We now want to show that $Z_j-\hat{Z}_j(Y)=Y_j-\hat{Y}_j(Y)$. Since we have $Y_j=Z_j-Z_D$ and
\begin{equation}
    Y_j-\hat{Y}_j=Z_j-Z_D-\widehat{(Z_j-Z_D)}=Z_j-\hat{Z}_j-Z_D+\hat{Z}_D,
\end{equation}
all we have to show is that $\hat{Z}_D=Z_D$. We have
\begin{multline}
    \hat{Z}_D(Y)=\mathrm{cov}\left(\log\frac{X_D}{g(X)},\log\frac{X}{X_D}\right)\mathbf{\Sigma}^{-1}Y=\mathrm{cov}\left(-\log\frac{g(X)}{X_D},\log\frac{X}{X_D}\right)\mathbf{\Sigma}^{-1}Y\\
    =\mathrm{cov}\left(-\frac{1}{D}\sum_{i=1}^{D-1}Y_i,Y\right)\mathbf{\Sigma}^{-1}Y=-\frac{1}{D}\left(\sum_{i=1}^{D-1}\sigma_{il}\right)^T_{l=1,\dots,D-1}\mathbf{\Sigma}^{-1}Y=-\frac{1}{D}\left(\sum_{l=1}^{D-1}\sum_{i=1}^{D-1}\sigma_{il}\sigma_{lk}^{(-1)}\right)^T_{k=1,\dots,D-1}Y\\
    =-\frac{1}{D}\left(\sum_{i=1}^{D-1}\delta_{ik}\right)^T_{k=1,\dots,D-1}Y=-\frac{1}{D}\sum_{k=1}^{D-1}Y_k=-\frac{1}{D}\sum_{k=1}^{D-1}\log\frac{X_k}{X_D}=\log\frac{X_D}{g(X)}=Z_D.
\end{multline}
Note that for the first equality of the third line to be true, the geometric mean has to run over the same set as the variables we are partialling on. Here, the fact that we include the reference $X_D$ in the geometric mean only plays a role with respect to the size of the prefactor $1/D$, as $\log(X_D/X_D)=0$. Together with the independence from the transformation used in the explanatory ratios shown in (\ref{YisZ}), we conclude that $Y_j-\hat{Y}(Y)=Z_j-\hat{Z}(Z)$.

\subsection*{Pseudoinverse of $\mathbf{\Gamma}$ under permutations of parts in $\mathbf{\Sigma}$}
Here we show that a permutation of parts in $Y$, the variables having covariance $\mathbf{\Sigma}$, corresponds to a simple permutation of the rows in the pseudoinverse $\mathbf{\Gamma}^{-}$ of the covariance of the corresponding clr-transformed random vector $Z$. Permuting rows in $\mathbf{\Gamma}^{-}$ gives
\begin{multline}
   \mathbf{P}\mathbf{\Gamma}^{-}\mathbf{P}^T=\mathbf{P}\mathbf{F}^T\mathbf{\Sigma}^{-1}\mathbf{F}\mathbf{P}^T=\mathbf{P}\mathbf{F}^T\left(\mathbf{Q}_P^T(\mathbf{Q}_P^T)^{-1}\right)\mathbf{\Sigma}^{-1}\left(\mathbf{Q}_P^{-1}\mathbf{Q}_P\right)\mathbf{F}\mathbf{P}^T\\
   =\mathbf{P}\mathbf{F}^T\mathbf{Q}_P^T\left(\mathbf{Q}_P\mathbf{\Sigma}\mathbf{Q}_P^T\right)^{-1}\mathbf{Q}_P\mathbf{F}\mathbf{P}^T
    =\mathbf{P}\mathbf{F}^T\mathbf{Q}_P^T\mathbf{\Sigma}^{-1}_P\mathbf{Q}_P\mathbf{F}\mathbf{P}^T,\label{pseudoperm}
\end{multline}
where we applied (\ref{pseudo}), inserted two identity matrices, and used the identity for permutation of parts in $\mathbf{\Sigma}$ (\ref{v}). We now observe that
\begin{equation}
    \mathbf{F}^T\mathbf{Q}_P^T=\mathbf{F}^T\left(\mathbf{F}\mathbf{P}\mathbf{F}^T\mathbf{H}^{-1}\right)^T=\mathbf{F}^T(\mathbf{H}^{-1})^T\left(\mathbf{F}\mathbf{P}\mathbf{F}^T\right)^T=\mathbf{F}^T\mathbf{H}^{-1}\mathbf{F}\mathbf{P}^T\mathbf{F}^T,\label{knutzel}
\end{equation}
where we used the symmetry of $\mathbf{H}$ for the last equality. This expression contains another auxiliary matrix $\mathbf{G}$ described in (Aitchsion, 1986), which is defined as the $D\times D$ matrix obtained by subtracting from the identity matrix the matrix consisting of elements 1/D. By Properties F1 and G3 in the appendix of Aitchison's book and Properties 5.4 there, we have
\begin{eqnarray}
    \mathbf{G}&=&\mathbf{F}^T\mathbf{H}^{-1}\mathbf{F},\\
    \mathbf{P}\mathbf{G}\mathbf{P}^T&=&\mathbf{G},\\
    \mathbf{F}\mathbf{G}&=&\mathbf{F}.
\end{eqnarray}
With (\ref{knutzel}) and these properties, (\ref{pseudoperm}) simplifies to
\begin{equation}
    \mathbf{P}\mathbf{F}^T\mathbf{Q}_P^T\mathbf{\Sigma}^{-1}_P\mathbf{Q}_P\mathbf{F}\mathbf{P}^T=\mathbf{P}\mathbf{G}\mathbf{P}^T\mathbf{F}^T\mathbf{\Sigma}^{-1}_P\mathbf{F}\mathbf{P}\mathbf{G}\mathbf{P}^T=\mathbf{G}\mathbf{F}^T\mathbf{\Sigma}^{-1}_P\mathbf{F}\mathbf{G}=\mathbf{F}^T\mathbf{\Sigma}^{-1}_P\mathbf{F}=\mathbf{\Gamma}^{-}_P.
\end{equation}
The last identity follows from (\ref{pseudo}) again. We have shown that $\mathbf{P}\mathbf{\Gamma}^{-}\mathbf{P}^T=\mathbf{\Gamma}^{-}_P$ through permutation of parts in $\mathbf{\Sigma}$. In Property 5.2 (b) in (Aitchison, 1986) the corresponding identity is stated for $\mathbf{\Gamma}_P$ instead of its pseudoinverse.

\subsection*{Permutation test and local false discovery rates}
For the general framework, we follow the procedure outlined in (Hastie et al, 2001). As permuting the residuals seems to lead to a test that is too lenient, we opted for permuting the samples in each column of the original data matrix before evaluating and inverting the covariance matrix. A false-discovery rate (FDR) estimate is obtained for a given cut-off $C$ from dividing the average number of pairs with randomized partial correlations above the cut-off $N_\mathrm{ra}(C)$ by the number of pairs with true partial correlations above the cut-off $N_\mathrm{ob}(C)$. The number of randomizations was 10,000. All FDRs for cut-offs $C$ in steps of 0.001 were evaluated and a $q$-value for a given partial correlation value $r$ was determined by taking the minimum over the FDRs corresponding to $C\le r$. For negative $r$, the FDRs were determined separately on the other tail of the distribution.

\end{document}